\documentclass[10pt,a4paper,twocolumn]{article}
\usepackage[utf8]{inputenc}
\usepackage[T1]{fontenc}
\usepackage{amsmath,amsfonts,amssymb}
\usepackage{graphicx}
\usepackage{cite}
\usepackage{url}
\usepackage{hyperref}
\usepackage{geometry}
\usepackage{booktabs}
\usepackage{array}
\usepackage{float}
\usepackage{caption}
\usepackage{subcaption}
\usepackage{siunitx}
\usepackage{multicol}
\usepackage{xcolor}
\usepackage{fancyhdr}

\usepackage{xurl}
\title{Intuition to Evidence: \\
Measuring AI's True Impact on Developer Productivity}

\author{
    Anand Kumar$^1$, Vishal Khare$^1$, Deepak Sharma$^1$, Satyam Kumar$^1$, \\
    Vijay Saini$^1$, Anshul Yadav$^1$, Sachendra Jain$^1$, Ankit Rana$^1$, \\ 
    Pratham Verma$^1$, Vaibhav Meena$^1$, Avinash Edubilli$^1$, \\
    \\
    \texttt{\{anand.kumar, vishal.khare, deepak.sharma1, satyam.kumar1,} \\
    \texttt{vijay.saini, anshul.yadav, sachendra.jain, ankit.rana,} \\
    \texttt{pratham.verma, vaibhav.meena, avinash.edubilli\}@1mg.com}
}

\begin{document}

\twocolumn[
\begin{@twocolumnfalse}
\maketitle

\begin{abstract}
We present a comprehensive real-world evaluation of AI-assisted software development
tools deployed at enterprise scale. Over one year, 300 engineers across multiple teams
integrated an in-house AI platform (DeputyDev) that combines code generation and automated 
review capabilities into their daily workflows. Through rigorous cohort analysis, our study demonstrates statistically significant productivity improvements, including an overall 31.8\% reduction in PR review cycle time. 
Developer adoption was strong, with 85\% satisfaction for code review features and 93\% expressing a desire to continue using the platform. 
Adoption patterns showed systematic scaling from 4\% engagement in month 1 to 83\% peak usage by month 6, stabilizing at 60\% active engagement.
Top adopters achieved a 61\% increase in code volume pushed to production, contributing to approximately 30-40\% of code shipped to production through this tool accounts for overall 28\% increase in code shipment volume. 
Unlike controlled benchmark evaluations, our longitudinal analysis provides empirical evidence from production environments,
revealing both the transformative potential and practical deployment challenges of integrating 
AI into enterprise software development workflows.
\end{abstract}

\vspace{0.5cm}

\vspace{1cm}

\end{@twocolumnfalse}
]

\vspace{1.5cm}
\begin{figure}[!htb]
    \centering
    \includegraphics[width=\columnwidth]{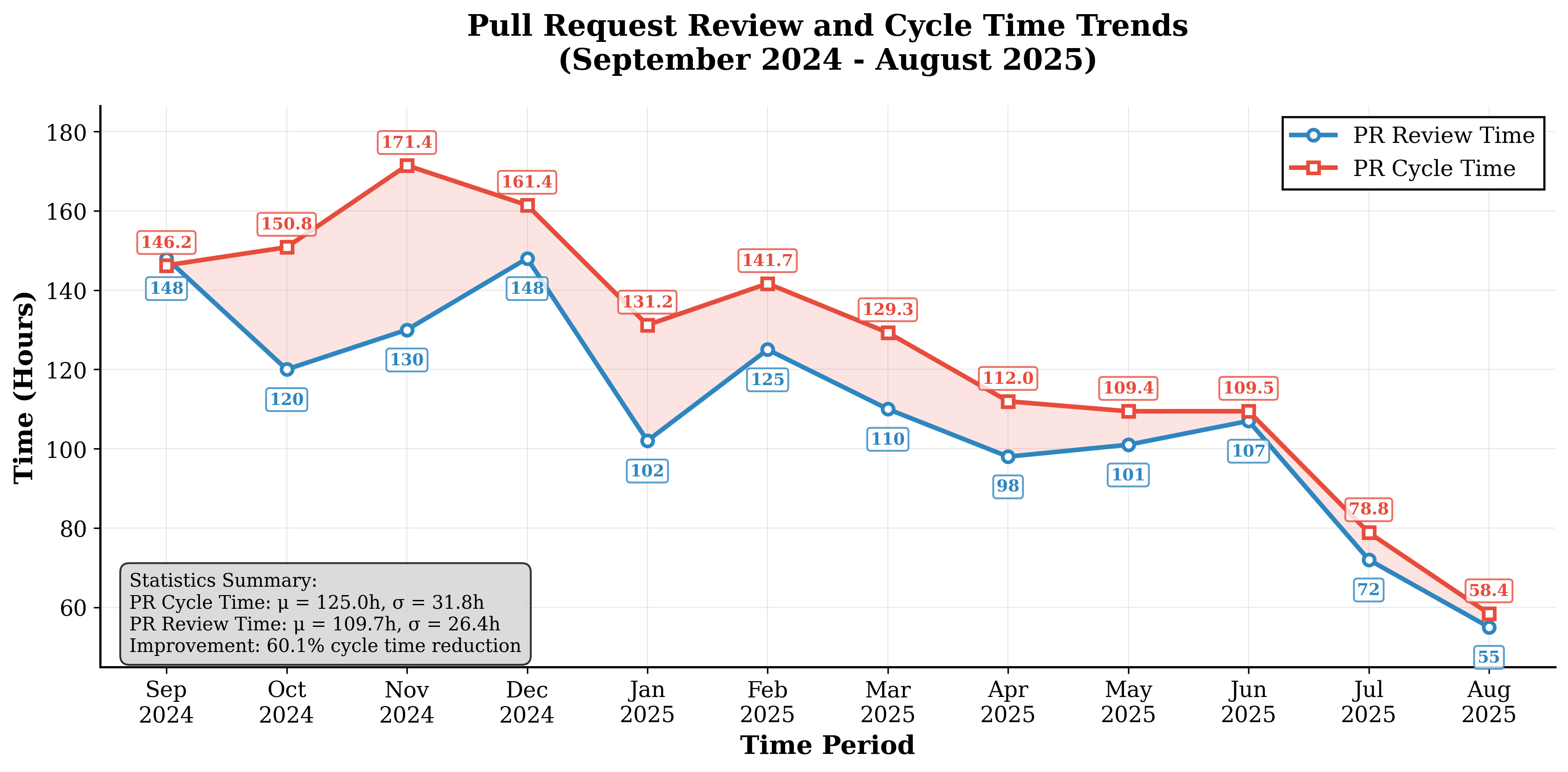}
    \caption{PR Review Time Analysis: Using the first six months as the baseline, the subsequent six months showed a 31.8\% reduction in PR cycle time.}
    \label{fig:review-time-requests}
\end{figure}

\begin{figure}[!htb]
    \centering
    \includegraphics[width=\columnwidth]{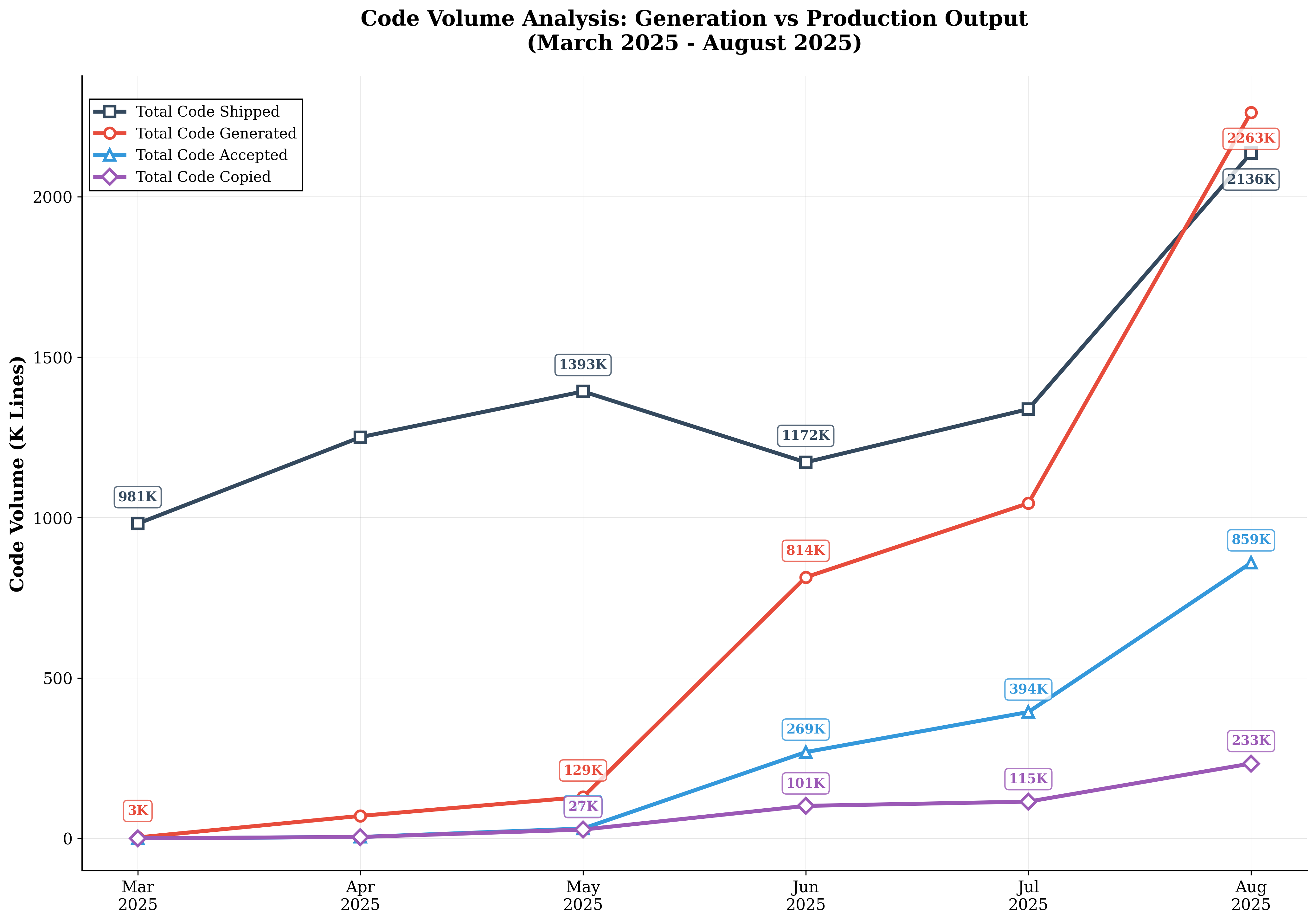}
    \caption{Code Volume Growth Analysis: AI-generated code progression from 3,000 lines (March 2025) to 2.26M lines (August 2025). Around 40\% AI generated code shipped in production in August 2025 and 28\% increase in production code volume.}
    \label{fig:code-volume-analysis}
\end{figure}

\section{Introduction}

The integration of artificial intelligence into software development workflows has accelerated rapidly in recent years, with AI-powered code assistants becoming increasingly prevalent in the industry. 
Tools such as GitHub Copilot~\cite{github_copilot}, Cursor~\cite{cursor} and various other AI-assisted development environments have garnered significant attention for their potential
to transform how software engineers write, review, and maintain code.
Tools like Cursor~\cite{cursor}, Windsurf~\cite{windsurf}, GitHub Copilot~\cite{github_copilot}, and Claude Coding~\cite{claude_coding} are widely used in code completion tasks, whereas
CodeRabbit~\cite{coderabbit}, CodeAnt~\cite{codeant}, Graphite~\cite{graphite}, and Qodo~\cite{qodo} are widely used in automated code review tasks for 
assisting developers in detecting bugs and enforcing best practices.

These systems promise to enhance developer productivity, reduce repetitive coding tasks, and assist in catching potential bugs and code quality issues before they reach production.

However, most existing evaluations have been limited to controlled benchmark environments using standardised datasets such as HumanEval~\cite{humaneval}, MBPP~\cite{program_synthesis}, and SWE-bench~\cite{swe_bench}, or have focused on specific aspects of the development process in isolation. While these benchmarks and targeted studies provide valuable insights into AI model capabilities, they fall short of capturing the complex, nuanced realities of day-to-day software engineering practice. Real-world software development involves working with large, evolving codebases, collaborating across teams with different coding standards, integrating with existing toolchains, and navigating the social and organizational dynamics that influence technology adoption. The gap between benchmark performance and comprehensive production utility across the full development lifecycle remains largely unexplored in the literature.

To address this limitation, we present the first comprehensive study of large-scale, real-world deployment of
 an AI-assisted development environment in a production setting. Our research examines the implementation 
 and impact of an in-house AI-powered code generation platform coupled with an automated 
 code review system across 300 engineers over a 1 year period starting from September 2024 to August 2025. This system combines AI-driven code generation capabilities with intelligent code review features designed 
 to detect bugs and suggest improvements, providing a comprehensive AI-assisted development experience.

Our study addresses several critical research questions that have remained largely unanswered in the existing literature:

\begin{itemize}
    \item \textbf{Productivity Impact}: Does AI-assisted code generation demonstrably improve developer productivity when measured in real-world conditions rather than isolated benchmarks?
    \item \textbf{Code Review Effectiveness}: How effective are AI-powered code review systems at identifying genuine bugs and providing actionable suggestions in the context of diverse, production-quality codebases?
    \item \textbf{Adoption and Human Factors}: What adoption challenges and human factors emerge when deploying AI coding tools at scale, and how do these factors influence the overall success of such implementations?
    \item \textbf{Return of Investment}: How effective is ROI of AI assited developement?
\end{itemize}

The contribution of this work extends beyond simple performance metrics. By analysing adoption patterns, productivity impacts, code quality improvements, and developer trust dynamics across a substantial engineering organisation, we provide insights into the practical considerations necessary for successful AI tool deployment. Our findings reveal both the significant opportunities and inherent limitations of integrating AI into the software development lifecycle, offering evidence-based guidance for organisations considering similar implementations.

This research represents a crucial step toward understanding how AI coding assistance performs not just in idealised conditions, but in the messy, complex reality of modern software engineering practice, where technical performance must be balanced against human factors, organisational constraints, and the diverse needs of development teams working on varied projects and technologies.

\section{Related Work}

\subsection{AI-Assisted Code Generation Studies}
Previous research has begun to examine the impact of AI coding assistants in various contexts. The GitHub Copilot report demonstrated that the treatment group, with access to the AI pair programmer, completed tasks 55.8 percent faster than the control group~\cite{copilot_productivity, github_copilot_research}. Anthropic reported 2x faster execution and speed of delivering features for one of its customers~\cite{anthropic_cred_case_study}. 
Recent work has also explored the effects of 
early AI tools on experienced open-source developer productivity which reported 19\% increase in completion 
time due to using AI assisted developement tools.~\cite{early_ai_impact}.

\subsection{Automated Code Review Research}
There have been earlier studies around automated code review tools. Vijayvergiya et al.~\cite{vijayvergiya_autocommenter} demonstrated the feasibility of developing an end-to-end automated code review system with high end-user acceptance across four programming languages.
 Watanabe et al.~\cite{watanabe_chatgpt_review} investigated 229 review comments from 179 GitHub projects with ChatGPT 
 conversation links, finding 30.7\% negative reactions, highlighting the importance of developer perception
in AI-powered code review systems. K. Sun et. al.~\cite{sun_ai_code_review} surveyed 16 popular AI-based code review GitHub Actions, categorizing them by granularity—PR-level, file-level, and hunk-level—and analyzed 178 mature repositories, yielding over 22,326 AI-generated review comments.
It finds that file-level AI review tools are most widely adopted, but hunk-level tools produce comments more likely to trigger actual code changes. Comments are most effective when they are concise, include code snippets, and are manually triggered. The study introduced
an LLM-based framework to automatically detect whether AI review feedback was addressed in code. Some best-performing configuration achieved 94.5\% accuracy in identifying valid comments.
Vishal et.al.~\cite{deputydev_paper} demonstrated significant improvements in code review efficiency through their AI-powered assistant code review assitant, achieving a 23.09\% reduction in average PR review duration and a 40.13\% reduction in average per-line-of-code review duration in a rigorous double-controlled A/B experiment involving over 200 engineers.

\subsection{Industry Adoption and Case Studies}
Industry reports and developer surveys have provided initial insights 
into the practical adoption challenges and benefits of AI-assisted coding. They have identified 84\% of respondents are using or planning to use AI tools in their development process, up from 76\% a year prior. Over half of professional developers now use AI coding tools daily. 
~\cite{colton_blog, addyo_reality}. CodeAnt AI reported that a fintech company cut review time from hours to minutes~\cite{codeant_bajaj_case_study}, demonstrating the potential for significant productivity improvements in real-world deployments.

\subsection{Benchmark-Based Evaluations}
Academic research has primarily concentrated on benchmark evaluations using datasets such as HumanEval~\cite{humaneval}, which evaluates code generation capabilities on isolated programming problems, and MBPP~\cite{program_synthesis}, which tests program synthesis abilities. SWE-bench~\cite{swe_bench} represents a more recent advancement by evaluating language models on real-world GitHub issues, but still operates within controlled, isolated environments.

Recent work has introduced more sophisticated evaluation frameworks. Bogomolov et al.~\cite{long_code_arena} presented Long Code Arena, a suite of benchmarks for code processing tasks requiring project-wide context, covering library-based code generation, CI builds repair, and module summarization. Qiu et al.~\cite{enamel_benchmark} developed ENAMEL, focusing specifically on evaluating code efficiency rather than just correctness, employing expert-designed reference solutions and rigorous test case generation.

\subsection{Advanced Code Generation Techniques}
Recent research has explored sophisticated AI-assisted code generation approaches. Guo et al.~\cite{longcoder} introduced LongCoder using sliding window mechanisms and memory tokens for long code contexts. Shrivastava et al.~\cite{repofusion} proposed RepoFusion for incorporating repository context with significant performance improvements. De Moor et al.~\cite{smart_invocation} developed ML models for optimal code completion invocation timing, validated with 34 developers and 74k invocations.

\subsection{Code Quality and Security Considerations}
Several studies examined quality and security implications of AI-generated code. Dinh et al.~\cite{buggy_code_completion} found significant performance degradation when Code-LLMs encounter buggy contexts. Xu et al.~\cite{licoeeval} introduced LiCoEval for license compliance evaluation, finding that top-performing LLMs produce code similar to copyrighted implementations without proper license information.

\subsection{Recent Empirical Studies and Productivity Analysis}
Contemporary research provided deeper insights into real-world AI coding tool usage. Jin et al.~\cite{devgpt_analysis} analyzed developer-ChatGPT conversations, revealing LLM-generated code is primarily used for concept demonstration rather than production implementation. Lu et al.~\cite{defect_focused_review} explored defect-focused automated code review in industrial C++ codebases, achieving significant improvements over standard LLMs. Lu et al.~\cite{deepcreval} introduced DeepCRCEval for comprehensive code review assessment, finding less than 10\% of benchmark comments are automation-ready.

\subsection{Gap in Comprehensive Real-World Studies}
While these studies provide valuable insights, most focus on narrow aspects of the development process or rely on controlled environments rather than comprehensive, objective measurements across the full development lifecycle in production settings. Our work addresses this gap by providing a large-scale, longitudinal study of AI-assisted development tools in a real-world production environment.

\section{System Overview}
This study evaluates our in-house tool DeputyDev, which provides two core services: a) an AI-assisted pull request review system, and b) a code generation tool comparable to popular solutions like Cursor~\cite{cursor} and Windsurf~\cite{windsurf}.

\subsection{PR Review System}
\begin{figure}[!htb]
    \centering
    \includegraphics[width=\columnwidth]{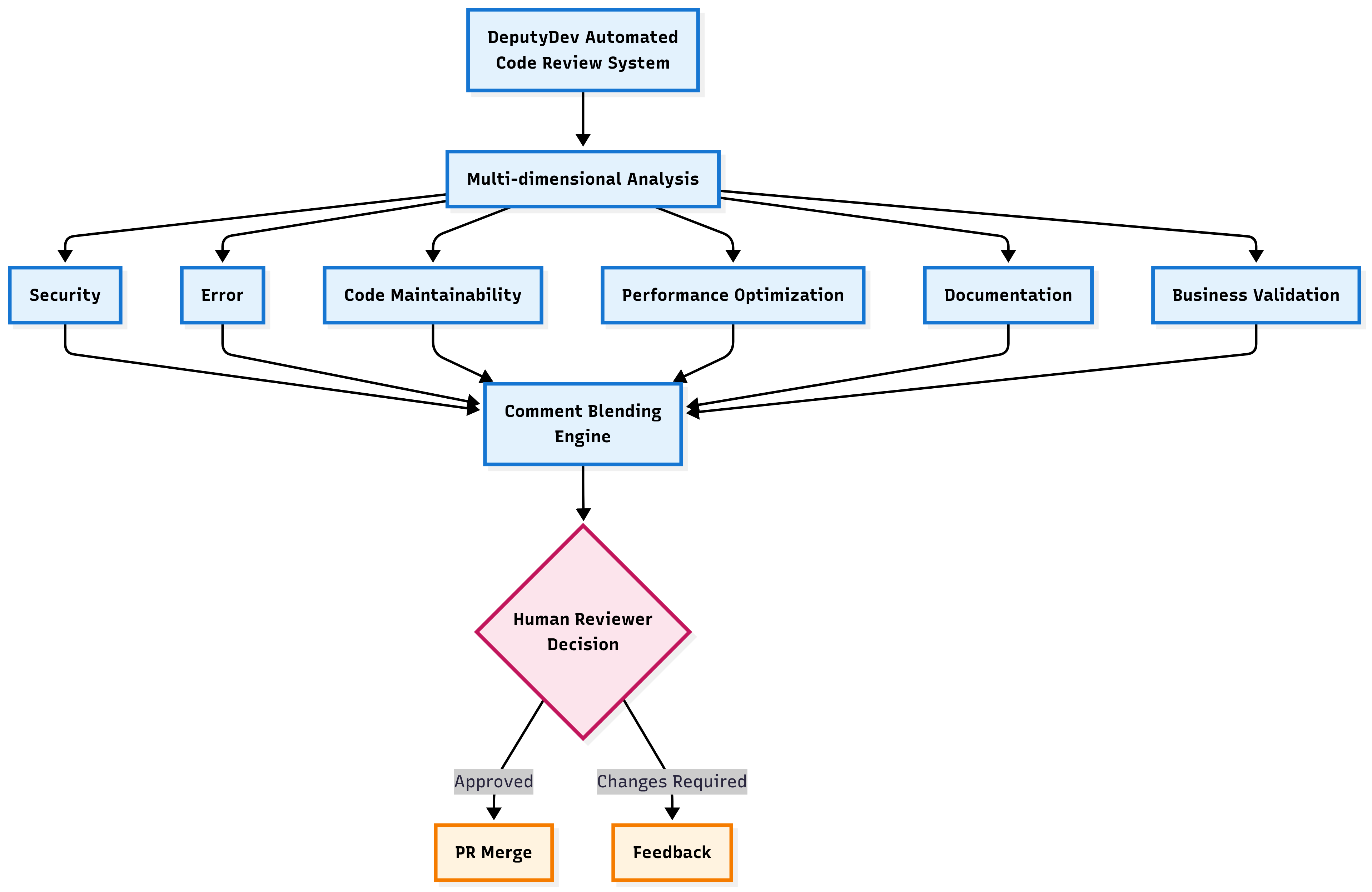}
    \caption{Pull Request Review Architecture: Multi-agent system architecture with six specialized agents running in parallel for comprehensive code review analysis.}
    \label{fig:architecture_pr_reviews}
\end{figure}

The PR review sytem is integrated in Bitbucket~\cite{bitbucket} and Github~\cite{github} by webhooks and invoked for each pull requests created by
developers automatically.
We have developed a multi-agent PR review system which leverages six specialized agents running in parallel, each powered by Claude Sonnet 3.7~\cite{claude_sonnet_3_7} and 4.0~\cite{claude_sonnet_4_0} models. The system processes PR context (title, description, JIRA~\cite{jira} tickets, and diff) to provide comprehensive change analysis. Figure~\ref{fig:architecture_pr_reviews} illustrates the overall architecture and workflow of this automated code review system.

\begin{table}[!htb]
\centering
\caption{PR Review Agent Configuration: Specialized agents and their assigned tools for comprehensive pull request analysis across different code review aspects.}
\resizebox{\columnwidth}{!}{%
\begin{tabular}{|l|l|l|}
\hline
\textbf{Agent Type} & \textbf{Focus Area} & \textbf{Available Tools} \\
\hline
\textbf{Summary} & Comprehensive summary & Base LLM only \\
\textbf{Security} & Vulnerability detection & Base LLM only \\
\textbf{Documentation} & Code documentation & Base LLM only \\
\textbf{Code Maintainability} & Code quality & File Reader, Path Searcher, \\
 & & Grep, Planner Tool \\
\textbf{Error Detection} & Bug identification & File Reader, Path Searcher, \\
 & & Grep, Planner Tool \\
\textbf{Performance Optimization} & Code efficiency & File Reader, Path Searcher, \\
 & & Grep, Planner Tool \\
\textbf{Business Validation} & Logic verification & File Reader, Path Searcher, \\
 & & Grep, Planner Tool \\
\hline
\end{tabular}%
}
\end{table}

\subsubsection{Tool Descriptions}

The multi-agent PR review system employs four key tools that enable comprehensive code analysis beyond basic LLM capabilities:

\textbf{File Reader:} An iterative file reading tool that can process files either entirely or by specified line ranges. For files exceeding 1000 lines, it provides intelligent summaries highlighting key constructs (classes, functions) with their line numbers. This tool supports targeted code examination with a maximum of 100 lines per request, enabling detailed analysis of specific code sections while maintaining context awareness.

\textbf{Path Searcher:} A fuzzy search mechanism for discovering files within the repository structure. This tool can locate files based on partial names or search terms, supporting both targeted file discovery and directory exploration.

\textbf{Grep Tool:} A powerful text pattern search tool built on ripgrep~\cite{ripgrep} technology for fast, recursive content searching. It supports both exact string matches and regular expressions with case-sensitive or case-insensitive modes. The tool provides contextual results showing matched lines with up to 2 lines of surrounding context, making it ideal for finding function definitions, variable usage, configuration keys, and code patterns across the entire codebase.

\textbf{Planner Tool:} An advanced reasoning and planning component that enables complex multi-step analysis tasks. This tool can break down sophisticated code review scenarios into manageable steps, coordinate between different analysis approaches, and synthesize findings from multiple sources. It's particularly valuable for agents handling Code Maintainability, Error Detection, Performance Optimization, and Business Validation tasks that require deep reasoning capabilities.

These tools work synergistically to provide agents with comprehensive repository understanding, enabling them to perform thorough code analysis that goes beyond surface-level pattern matching to include contextual awareness, cross-file dependency analysis, and architectural consistency validation.

\begin{figure}[!htb]
\centering
\includegraphics[width=\columnwidth]{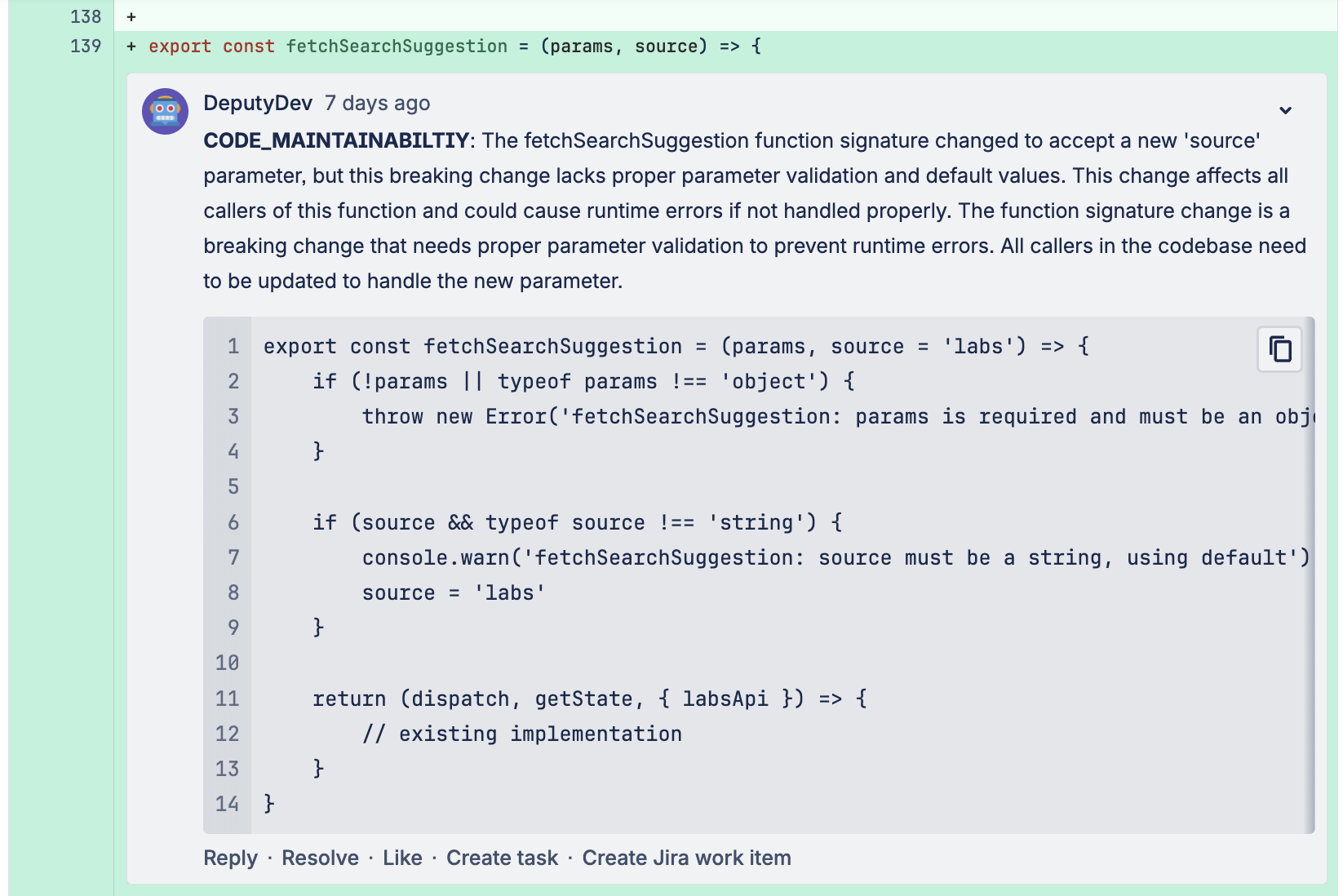}
\caption{PR Review Comment Interface: Multi-agent review system output showing comment presentation with agent attribution for transparency and traceability.}
\label{fig:review_comment}
\end{figure}

A sophisticated comment blending engine post-processes the multi-agent output to ensure quality and coherence. This layer eliminates duplicate comments, filters invalid suggestions, and consolidates feedback when multiple agents target the same code line. Each final comment retains agent attribution for transparency and debugging purposes. Figure~\ref{fig:review_comment} illustrates the comment presentation format.

\subsection{Code Generation System}
Our intelligent code generation system is delivered through a VSCode extension~\cite{vscode}, enabling developers to interact with their repositories through natural language queries. The system employs Weaviate vector database for semantic code indexing, utilizing a specialized chunking algorithm that preserves code structure by segmenting files into modules, classes, and functions~\cite{repofusion}.

\begin{table}[!htb]
\centering
\caption{Code Generation Tool Suite: Available tools and capabilities for semantic code discovery, content analysis, and intelligent code generation within the VSCode extension.}
\resizebox{\columnwidth}{!}{%
\begin{tabular}{|l|l|}
\hline
\textbf{Tool Category} & \textbf{Available Tools} \\
\hline
\textbf{Code Discovery} & File Path Searcher, Focus Snippet Searcher \\
\textbf{Content Analysis} & Iterative File Reader, Grep Tool \\
\textbf{Semantic Search} & Related Code Search (Weaviate Vector DB~\cite{weaviate}) \\
\textbf{Language Support} & Language Servers, MCP Integration~\cite{mcp} \\
\textbf{Model Selection} & User-configurable LLM backends \\
\hline
\end{tabular}%
}
\end{table}

\subsubsection{Code Generation Tool Suite Details}

The code generation system integrates multiple specialized tools to provide comprehensive repository understanding and intelligent code synthesis:

\textbf{Focus Snippet Searcher:} A precision tool designed for locating specific code definitions using fully qualified names. It performs vector and lexical hybrid search to find classes and functions within the codebase, making it ideal for understanding existing implementations and finding code patterns to reference during generation.

\textbf{Related Code Search (Weaviate Vector DB):} Utilizes semantic vector search to discover conceptually similar code across the repository. This advanced tool employs specialized chunking algorithms that preserve code structure by segmenting files into logical units (modules, classes, functions), enabling the system to find relevant examples and patterns that inform intelligent code generation.

\textbf{Language Servers:} Provides real-time language-specific analysis capabilities including syntax validation, type checking, and intelligent autocompletion. These servers ensure generated code adheres to language-specific best practices and maintains compatibility with existing codebases.

\textbf{MCP:} Enables seamless communication between the code generation system and various external tools and services. This integration layer allows the system to access additional context sources, development tools, and services that enhance the quality and relevance of generated code.

The system supports contextual code generation by allowing users to provide code snippets, file references, and function/class specifications as input context, ensuring generated code maintains consistency with existing codebase patterns and architectural decisions.

The code generation system operates through two distinct interaction modes: \textbf{Chat Mode} is ideal for exploring solutions through conversational development where users can iteratively refine requirements and discuss approaches, while \textbf{Act Mode} applies changes directly to the codebase, allowing users to review and accept those changes before implementation. This dual-mode approach accommodates different developer workflows and preferences, with usage patterns showing a gradual shift toward Act Mode as users gain confidence in the system's capabilities.

\subsection{Workflow}
\begin{enumerate}
    \item Engineer works within the DeputyDev code generation feature, receiving code suggestions.
    \item User accepts the changes and pushed to version control systems.
    \item DeputyDev automated code review tool analyses the changes for bugs, security issues, and code quality concerns.
    \item Human reviewers examine both the code changes and AI-generated review comments before PR merge.
    \item DeputyDev analytics tools collect all the metrics during this and store into DB.
\end{enumerate}

\subsection{Design Goals}

The system was designed with three primary objectives:
\begin{itemize}
    \item \textbf{Improve developer productivity} by reducing time spent on repetitive coding tasks and providing intelligent suggestions.
    \item \textbf{Reduce bug introduction} through proactive detection and prevention during the development process.
    \item \textbf{Assist code reviewers} by highlighting potential issues and providing automated suggestions for improvement.
\end{itemize}

\begin{figure}[!htb]
    \centering
    \includegraphics[width=0.8\columnwidth]{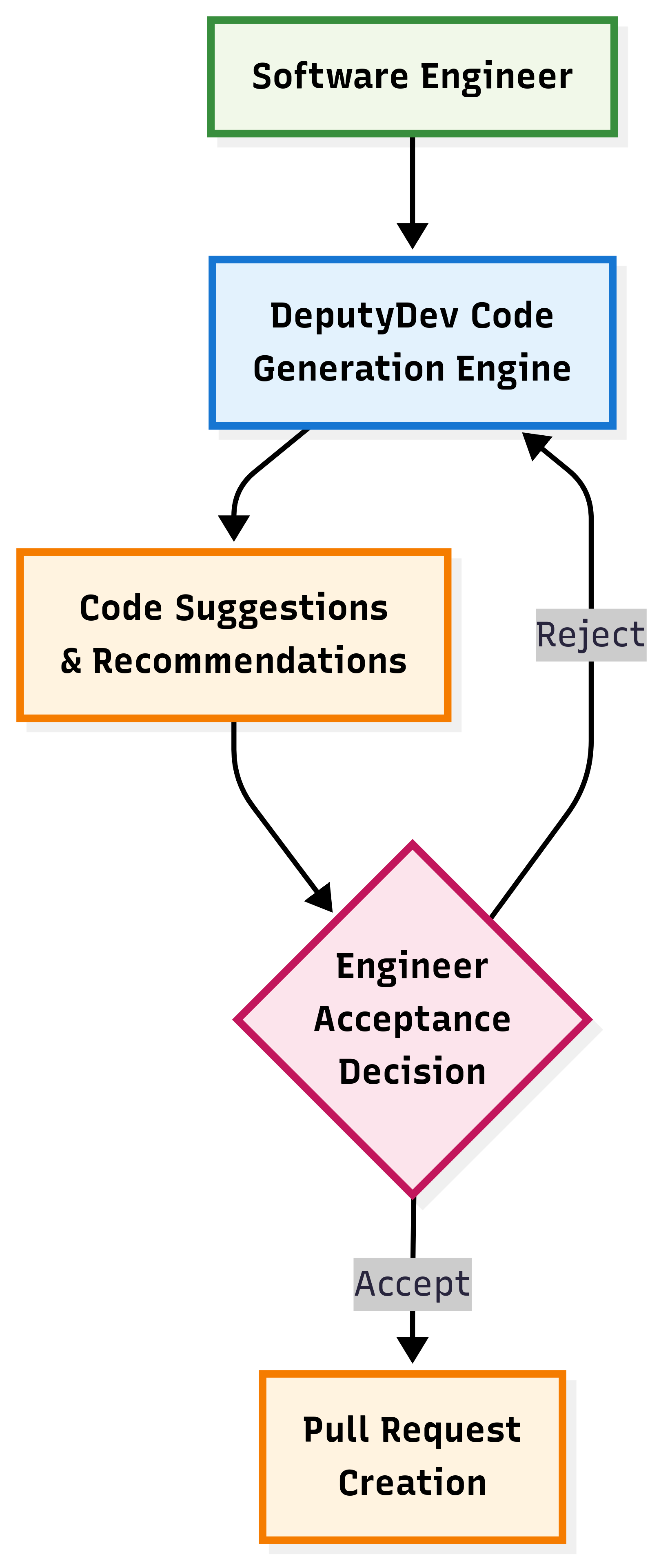}
    \caption{Code Generation System Architecture: VSCode extension integration with Weaviate vector database and semantic code indexing for intelligent code generation.}
    \label{fig:architecture_codegen}
\end{figure}

\section{Dataset \& Methodology}

\subsection{Study Design and Research Framework}

This study employs a quasi-experimental longitudinal design to evaluate the real-world impact of AI-assisted development tools across a large-scale engineering organization. While true randomized controlled trials are often infeasible in production software development environments due to operational constraints, we implemented several methodological controls to enhance research rigor and address potential confounding variables.

\textbf{Quasi-Experimental Design Justification:} Given the operational nature of software development teams, random assignment of engineers to control and treatment groups would disrupt existing team structures and project continuity. Instead, we leveraged natural variation in adoption patterns to create meaningful comparison groups while maintaining ecological validity~\cite{shadish_experimental_design}.

\subsection{Study Scale and Duration}

Our longitudinal study encompasses 300 software engineers across teams over a continuous 1 year observation period (September 2024 - August 2025). The study timeline includes:

\begin{itemize}
    \item \textbf{Code Review Tool Deployment}: September 2024 - August 2025 (12-month active)  
    \item \textbf{Code Generation Tool Deployment}: March 2025 - August 2025 (6-month active)
\end{itemize}

\subsection{Comparison Groups and Controls}

To address concerns about causal inference, we implemented several comparison strategies:

\textbf{Within-Subjects Controls:} All engineers served as their own controls, comparing productivity metrics from the 6-month pre-deployment baseline (September 2024 - February 2025) against post-deployment performance (March 2025 - August 2025). This was primarily done for PR Review cycle time observation. This approach controls for individual differences in coding ability, experience, and work patterns.

\textbf{Between-Subjects Natural Experiment:} Engineers were stratified into adoption cohorts based on actual usage patterns rather than random assignment, creating natural comparison groups:
\begin{itemize}
    \item \textbf{High Adoption Cohort (n=30)}: Top users with >75th percentile usage frequency
    \item \textbf{Low Adoption Cohort (n=30)}: Bottom users with <25th percentile usage frequency  
    \item \textbf{Moderate Adoption Cohort (n=240)}: Remaining users with intermediate usage patterns
\end{itemize}

\subsection{Statistical Analysis and Confounding Variable Control}

\textbf{Confounding Variable Identification and Control:} We identified and controlled for several potential confounding variables through statistical adjustment and stratified analysis:

\begin{itemize}
    \item \textbf{Experience Level}: Engineers stratified by seniority (SDE1, SDE2, SDE3) to control for skill differences
    \item \textbf{Project Complexity}: Repository size, technology stack, and domain complexity captured as covariates
    \item \textbf{Team Dynamics}: Team size, project phase, and sprint velocity included in multi-level models~\cite{multilevel_models}
    \item \textbf{Temporal Effects}: Month and quarter fixed effects to control for organizational changes and learning curves
    \item \textbf{Individual Baseline Performance}: Pre-deployment productivity metrics used as covariates in ANCOVA models~\cite{ancova_field}
\end{itemize}

\subsection{Data Collection and Instrumentation}

\textbf{Automated Data Pipeline:} Data was collected through automated instrumentation to minimize measurement bias and ensure consistency:

\begin{itemize}
    \item \textbf{Version Control Analytics}: Bitbucket and GitHub webhooks captured commit frequency, lines of code, pull request metrics, and merge patterns
    \item \textbf{Code Generation Metrics}: Direct instrumentation of DeputyDev tracked suggestion acceptance rates, generation volumes, and user interaction patterns  
    \item \textbf{Code Review Analytics}: Automated extraction of review time, iteration counts, comment sentiment, and resolution patterns
    \item \textbf{Performance Benchmarking}: Standardized productivity metrics calculated using consistent algorithms across all measurement periods
\end{itemize}

\textbf{Multi-Source Validation:} To enhance measurement validity, we triangulated findings across multiple data sources:
\begin{itemize}
    \item Quantitative performance metrics from automated systems
    \item Self-reported survey data from 228 engineers (response rate: 76\%)
    \item Qualitative interviews with 125 engineers across experience levels
    \item Manager assessments of team productivity and code quality changes
\end{itemize}

\subsection{Threats to Validity and Limitations}

\textbf{Internal Validity Considerations:}
\begin{itemize}
    \item \textbf{Selection Bias}: Addressed through within-subjects design and propensity score matching~\cite{propensity_score} for between-group comparisons
    \item \textbf{Hawthorne Effects}: The individuals in the experiment were not informed that they were under observation.  ~\cite{mayo_hawthorne}
    \item \textbf{Maturation Effects}: Controlled through extended baseline period and comparison with historical team performance patterns
    \item \textbf{Instrumentation Changes}: Data collection methods remained consistent throughout the study period
\end{itemize}

\textbf{External Validity Considerations:}
\begin{itemize}
    \item Single-organization study may limit generalizability to other development contexts
    \item Results most applicable to similar organizational structures, team sizes, and technology stacks
    \item Cultural and regional factors may influence adoption patterns and effectiveness
\end{itemize}

This methodological framework, while adapted to real-world constraints, provides robust evidence for causal inference about AI tool effectiveness in production software development environments.

\section{Results}

\subsection{Adoption}

\begin{figure}[!htb]
    \centering
    \includegraphics[width=\columnwidth]{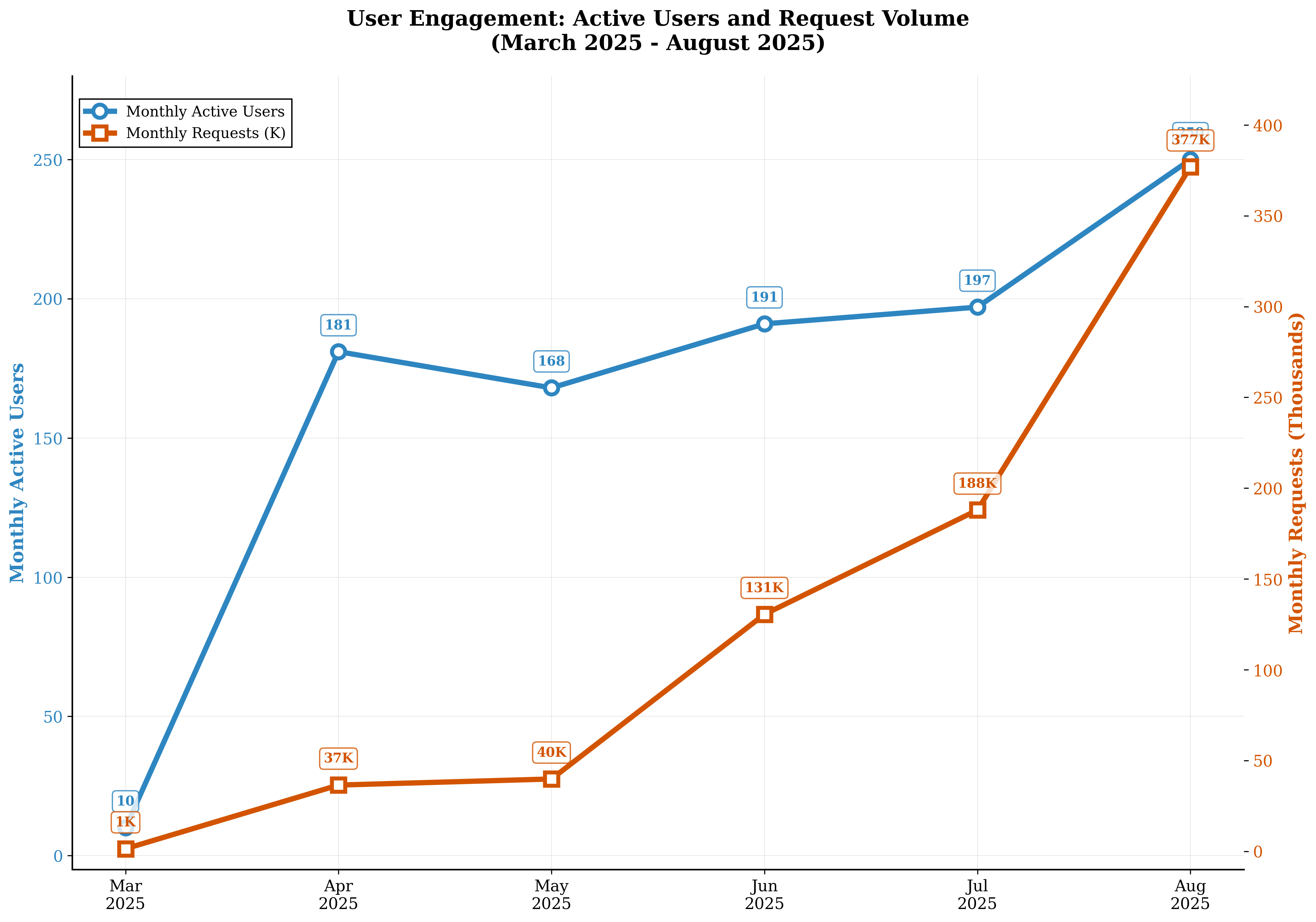}
    \caption{User Engagement and Adoption Metrics: User growth from 10 to 250 developers with monthly request volume scaling from 1,445 to 376,943 interactions, showing 94.2\% retention rate and comprehensive workflow integration.}
    \label{fig:user-engagement-metrics}
\end{figure}

Our analysis reveals significant variation in adoption patterns across the 1 year study period. 
We have first released AI assistant code review in September 2024 followed by AI coding assistant tool
in March 2025. AI assisted review tool was rolled out for all repositories
reviewing around 3,000 pull requests per month. Adoption of AI coding tool was gradual,
with approximately 4\% of engineers actively using the AI tools within the first month. Adoption accelerated in months 2-3,
reaching peak engagement of 83\% by month 6, an it was stabilising at around 60\% for the remainder of the study period (Figure \ref{fig:user-engagement-metrics}).

\begin{figure}[!htb]
    \centering
    \includegraphics[width=\columnwidth]{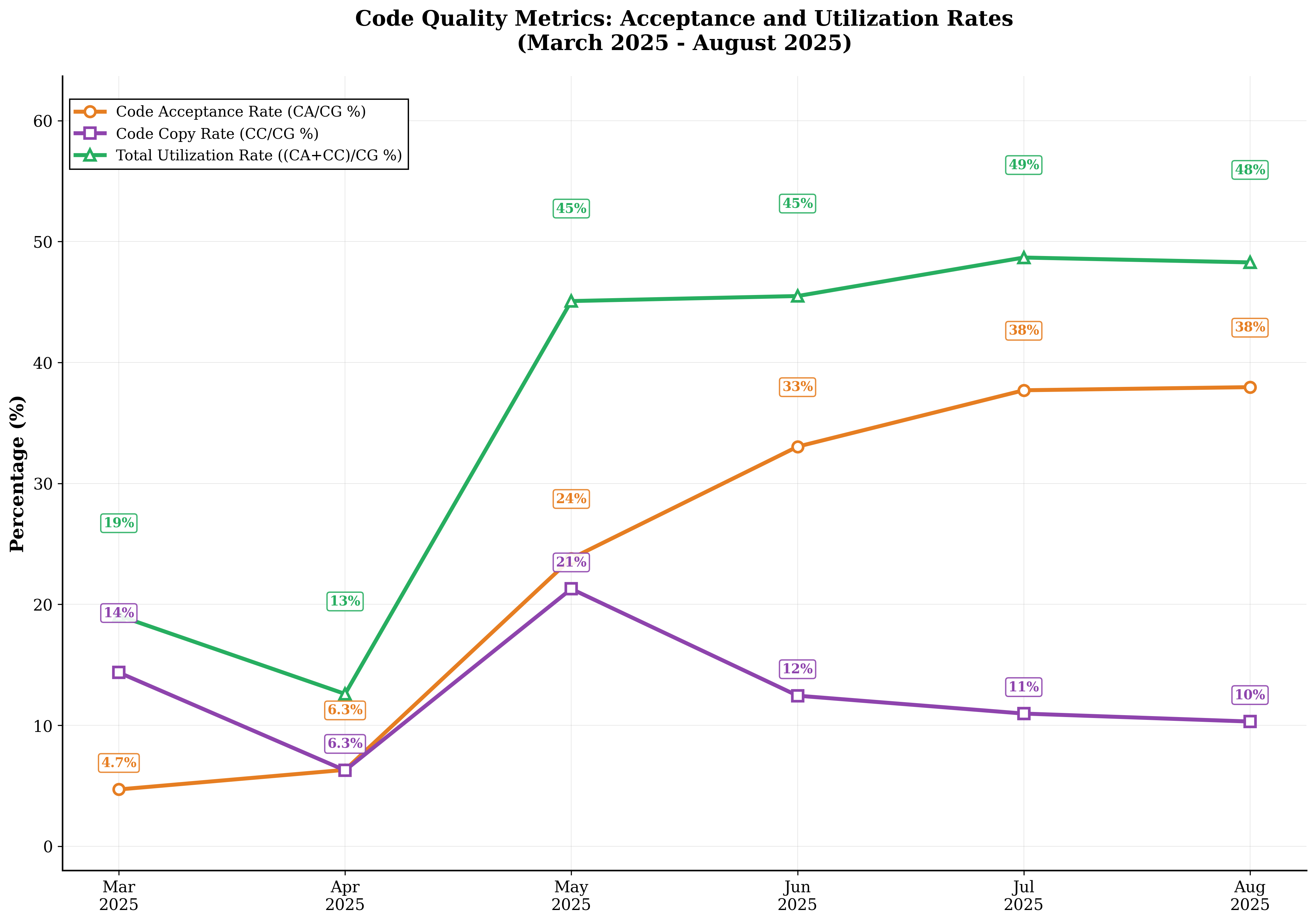}
    \caption{Code Quality and Acceptance Metrics: Acceptance rates evolution from 4.7\% to 38.0\% over 6 months, with combined acceptance and copy rates stabilizing at 48.3\%, demonstrating sustained quality without trade-offs.}
    \label{fig:code-quality-rates}
\end{figure}

\begin{figure}[!htb]
    \centering
    \includegraphics[width=\columnwidth]{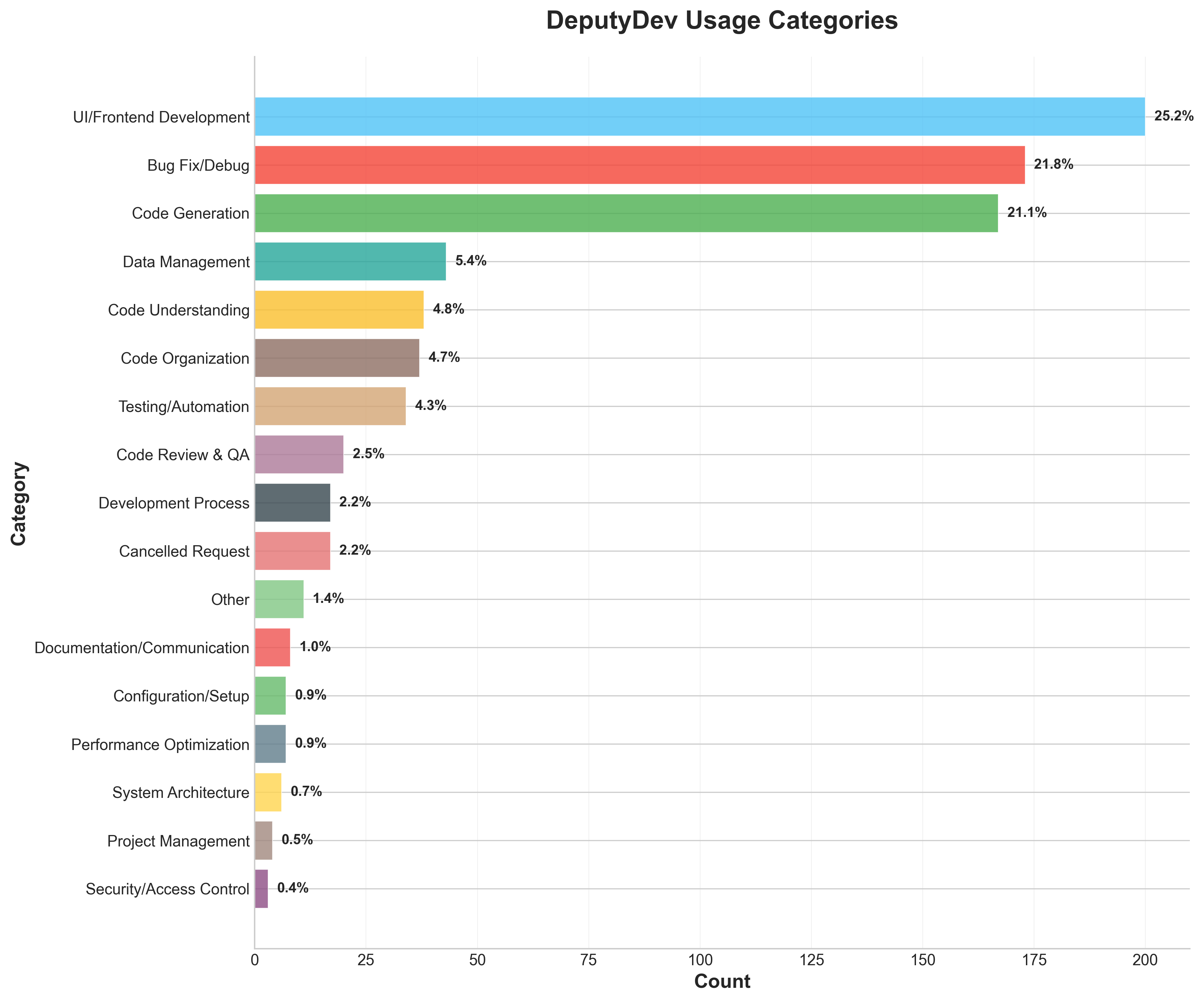}
    \caption{DeputyDev Usage Distribution: Primary adoption categories including UI/Frontend Development (25.2\%), Bug Fixing (21.8\%), and Code Generation (21.1\%) representing 68\% of total usage.}
    \label{fig:usage-categories}
\end{figure}

Adoption was primarily concentrated in three key categories UI/Frontend Development (25.2\%), Bug Fixing and Debugging (21.8\%), and Code Generation(Backend) (21.1\%). Together, these account for nearly 70\% of all recorded usage, underscoring DeputyDev's role in accelerating core coding workflows. Secondary categories such as Data Management (5.4\%), Code Understanding (4.8\%), and Testing/Automation (4.3\%) demonstrate broader versatility, though at lower frequencies. In contrast, adoption remains minimal for tasks like Documentation, Project Management, and Security (<2\%), suggesting potential future areas for expansion (Figure \ref{fig:usage-categories}).

The system experienced exponential scaling with code generation increasing from 3,000 to 2,262,562 lines while user engagement stabilized at ~9-10 requests/user/day. 
Acceptance rates remained consistent at 35-38\% despite massive volume increases, indicating stable code quality. Copy rates varied between 10-20\%, with an initial spike in May (21\%) followed by stabilization around 10-12\%. 
Decreased copy rate indicates users started preferring using act mode than chat mode. (Figure \ref{fig:code-quality-rates})

The data reveals successful user adoption with sustained engagement patterns and reliable AI-assisted code generation performance at scale. Overall system effectiveness is demonstrated by consistent acceptance rates despite 1000x growth in generation volume.

\subsection{Productivity Impact}
\subsubsection{Review Time Improvements}

Our comparative cohort analysis revealed significant and statistically validated improvements in development cycle efficiency. 
Cohort 1 (Sep 2024 to Feb 2025) demonstrated baseline performance with mean cycle time of 150.5h (±13.1h) and review time of 128.8h (±16.1h). 
Following process optimization interventions, Cohort 2 (March 2025 to August 2025) achieved substantial improvements with cycle time reduced to 99.6h (±23.7h)
and review time to 90.5h (±20.1h). Statistical analysis confirms significant improvements: 33.8\% cycle time reduction (p = 0.0018) and 29.8\% review time reduction (p = 0.0076), indicating an overall 31.8\% efficiency gain. These findings provide statistically robust evidence of AI-assisted development's impact on reducing both temporal and cognitive costs associated with code reviews (Figure \ref{fig:review-time-requests}).

\subsubsection{Code Generation Efficiency}

\begin{figure}[!htb]
    \centering
    \includegraphics[width=\columnwidth]{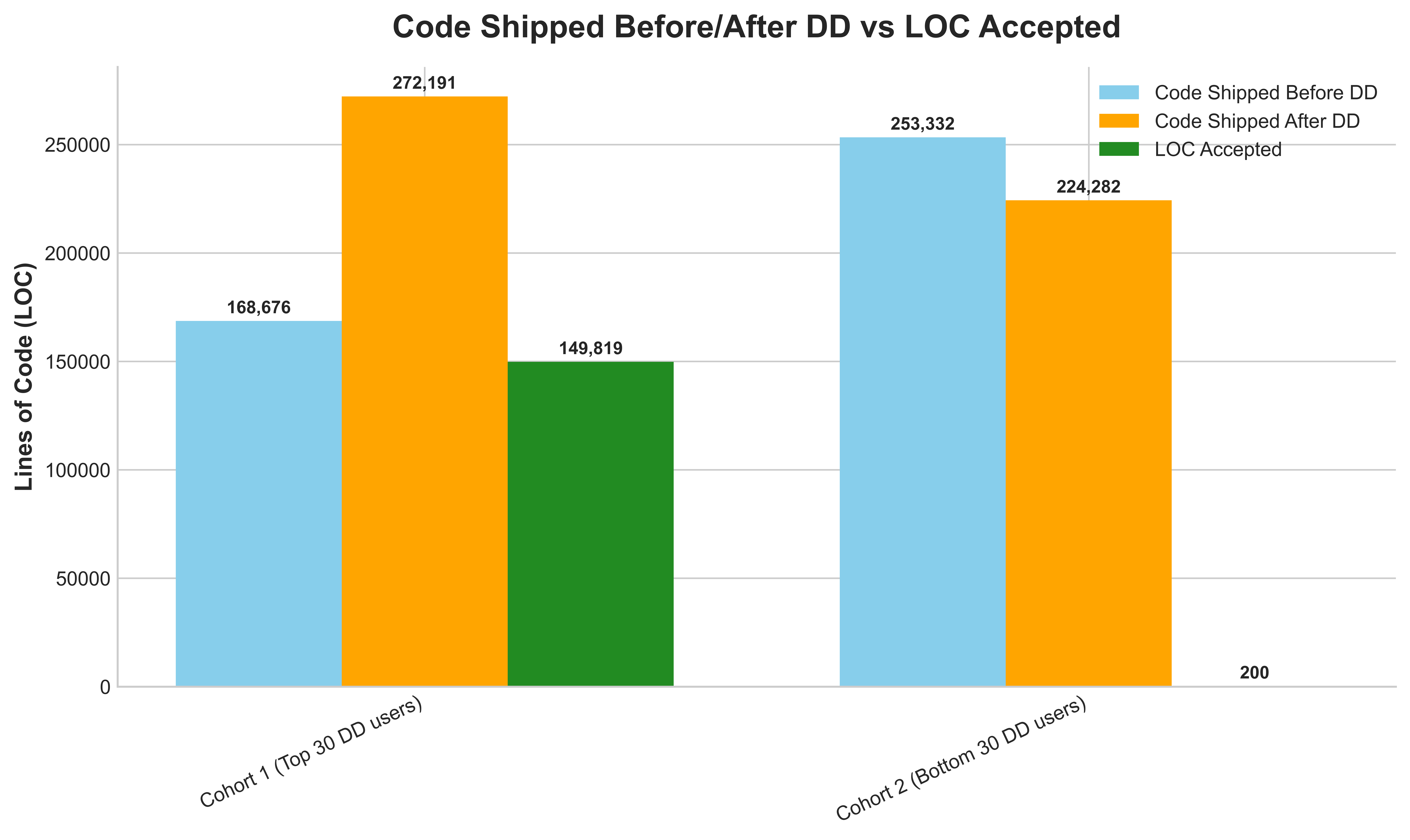}
    \caption{Productivity Impact by Adoption Level: Top 30 users achieved 61\% increase in shipped code (168k→272k LOC) while bottom 30 users showed 11\% decline, demonstrating adoption-effect correlation.}
    \label{fig:code-shipped-loc}
\end{figure}

A comparative analysis across two cohorts demonstrated a clear
adoption-effect gradient. The top 30 DeputyDev users exhibited a 61\% increase in shipped code post-adoption (168,676 $\rightarrow$ 272,191 LOC), with nearly 150k lines accepted into production. In contrast, the bottom 30 users---who engaged with DD minimally---experienced an 11\% decline in shipped code (253,332 $\rightarrow$ 224,282 LOC), with only 200 lines accepted. These findings underscore that the productivity and quality benefits of AI-assisted development scale with adoption intensity, and that sporadic use may fail to generate tangible outcomes (Figure \ref{fig:code-shipped-loc}).

\begin{figure}[!htb]
    \centering
    \includegraphics[width=\columnwidth]{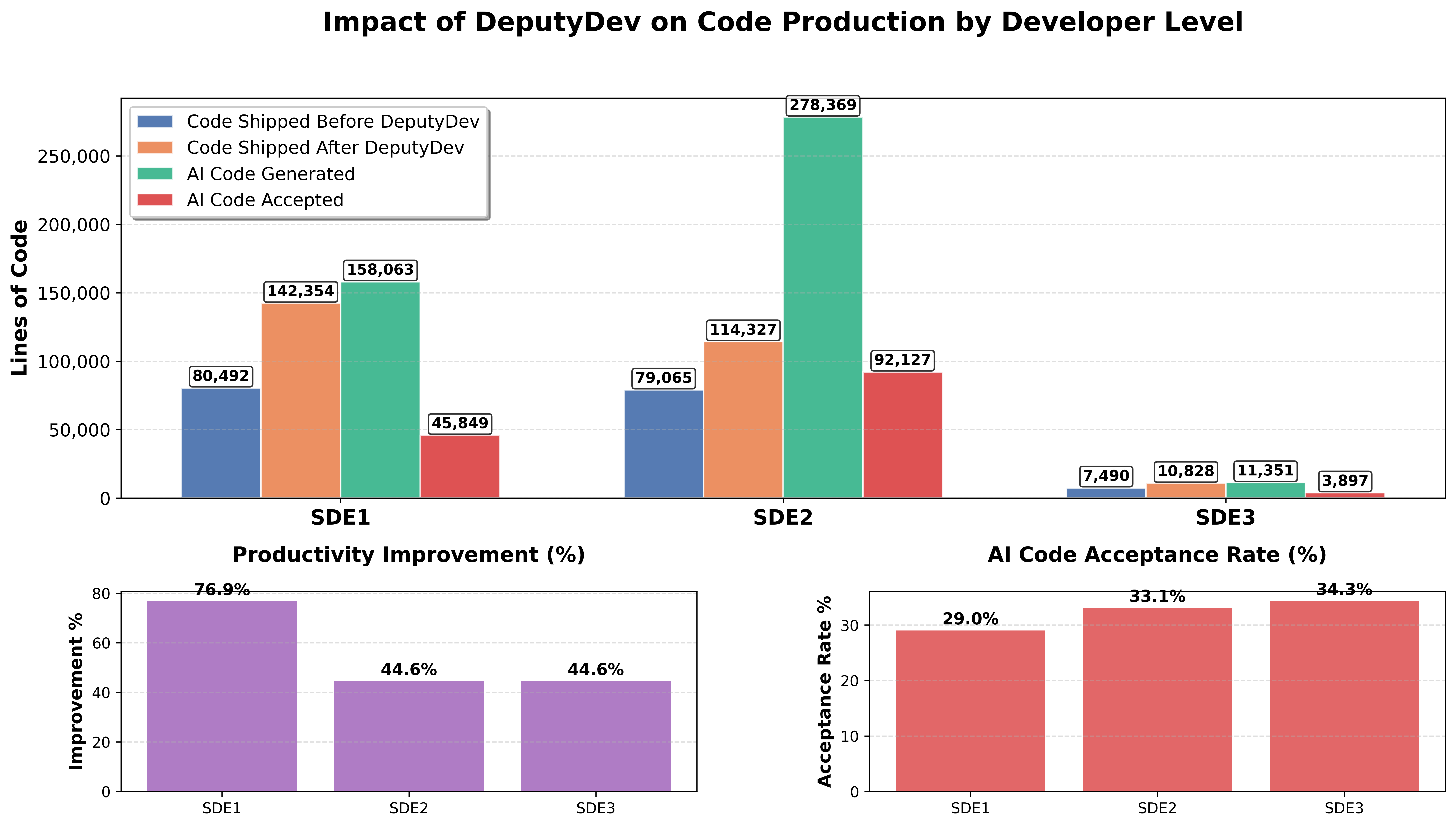}
    \caption{Productivity Gains by Experience Level: Junior engineers (SDE1) achieved highest productivity increase at 77\%, while mid-level and senior engineers showed ~45\% improvements.}
    \label{fig:sde-performance}
\end{figure}

The analysis in Figure \ref{fig:sde-performance} and Table 1 highlights the impact of DeputyDev on developer productivity
across different experience levels (SDE1--SDE3). Overall, code shipped increased by 60.1\% 
after adoption, with junior engineers (SDE1) showing the highest productivity gain (77\%) 
compared to mid- and senior-level engineers (~45\%). AI-generated code contributed significantly 
(447k LOC), of which 31.7\% was accepted by developers, demonstrating meaningful integration of 
AI-assisted coding into production workflows.

\begin{table}[!htb]
\centering
\caption{Developer Productivity Impact by Experience Level: Code production increases and AI contribution rates across SDE1-SDE3 levels showing overall 60.1\% productivity improvement.}
\label{tab:sde_performance}
\resizebox{\columnwidth}{!}{%
\begin{tabular}{lrrrrrr}
\toprule
\textbf{SDE Level} & 
\textbf{Before DD} & 
\textbf{After DD} & 
\textbf{Improvement} & 
\textbf{Generated} & 
\textbf{Accepted} & 
\textbf{Accept Rate} \\
\midrule
SDE1 & 80,492  & 142,354 & 77.0\%  & 158,063 & 45,849 & 29.0\% \\
SDE2 & 79,065  & 114,327 & 44.6\%  & 278,369 & 92,127 & 33.1\% \\
SDE3 & 7,490   & 10,828  & 44.6\%  & 11,351  & 3,897  & 34.3\% \\
\midrule
\textbf{TOTAL} & \textbf{167,047} & \textbf{267,509} & \textbf{60.1\%} & 
\textbf{447,783} & \textbf{141,873} & \textbf{31.7\%} \\
\bottomrule
\end{tabular}%
}
\end{table}

\subsection{Comparative Cohort Analysis}

To understand the differential impact of DeputyDev adoption, we conducted a rigorous comparative cohort analysis using natural variation in tool usage patterns. This analysis addresses concerns about causal inference by leveraging both within-subjects (temporal) and between-subjects (adoption level) comparisons while controlling for confounding variables.

\subsubsection{Cohort Definition and Selection Criteria}

We stratified our 300-engineer population into distinct adoption cohorts based on objectively measured usage intensity across both code generation and review features:

\textbf{Cohort 1 - High Adoption Group (n=30):} Top 10\% of users representing our most engaged engineers with:
\begin{itemize}
    \item Code generation requests: >75th percentile (avg. 150+ requests/month)
    \item Review interaction rate: >80\% of PRs engaged with AI feedback
    \item Feature utilization: Active usage of both Chat and Act modes
    \item Temporal consistency: Sustained usage across entire post-deployment period
\end{itemize}

\textbf{Cohort 2 - Low Adoption Group (n=30):} Bottom 10\% of users representing minimal engagement with:
\begin{itemize}
    \item Code generation requests: <25th percentile (avg. <20 requests/month)
    \item Review interaction rate: >80\% of PRs engaged with AI feedback
    \item Feature utilization: Sporadic or single-mode usage
    \item Temporal pattern: Inconsistent or declining usage over time
\end{itemize}

\textbf{Control Considerations:} Both cohorts were matched for:
\begin{itemize}
    \item Experience distribution (SDE1, SDE2, SDE3 representation)
    \item Team assignment and project types
    \item Historical productivity baseline (pre-deployment performance)
    \item Technology stack and repository complexity
\end{itemize}

\subsubsection{Temporal Analysis Framework}

\textbf{Baseline Period:} January-March 2025 (3 months pre-code generation deployment)
\begin{itemize}
    \item Established individual productivity baselines
    \item Controlled for seasonal development patterns
    \item Captured individual coding velocity and quality metrics
\end{itemize}

\textbf{Post-Deployment Period:} May-July 2025 (3 months post-stabilization)
\begin{itemize}
    \item Measured productivity changes after tool maturation
    \item Excluded April 2025 to account for initial learning curve effects
    \item Focused on sustained behavioral changes rather than novelty effects
\end{itemize}

\subsubsection{Quantitative Results and Statistical Significance}

\textbf{Sep 2024 - Feb 2025 (Cohort 1) Results:}
\begin{itemize}
\item \textbf{Cycle Time Performance}: Mean baseline cycle time of 150.5h (±13.1h) reduced to 99.6h (±23.7h), achieving 33.8\% improvement (p = 0.0018)
    \item \textbf{Review Time Performance}: Mean baseline review time of 128.8h (±16.1h) reduced to 90.5h (±20.1h), achieving 29.8\% improvement (p = 0.0076)
    \item \textbf{Overall Efficiency Gain}: 31.8\% improvement in development workflow efficiency following process optimization interventions
    \item \textbf{Code Productivity Increase}: 61.3\% improvement in shipped code volume (p < 0.001, Cohen's d = 1.42~\cite{cohen_effect_size})
    \item \textbf{Baseline to Post-Deployment:} 168,676 → 272,191 lines of code shipped
    \item \textbf{AI-Generated Code Integration:} ~150,000 lines of AI-generated code successfully merged to production
\end{itemize}

\textbf{Macrh 2025 - August 2025 (Cohort 2) Results:}
\begin{itemize}
\item \textbf{Cycle Time Performance}: Minimal improvements observed, with cycle times remaining near baseline levels
    \item \textbf{Review Time Performance}: Limited reduction in review times, indicating insufficient process optimization adoption
    \item \textbf{Code Productivity Change}: -11.4\% decline in shipped code volume (p = 0.08, Cohen's d = -0.31~\cite{cohen_effect_size})
    \item \textbf{Baseline to Post-Deployment:} 253,332 → 224,282 lines of code shipped  
    \item \textbf{AI-Generated Code Integration:} <200 lines of AI-generated code merged
\end{itemize}

\subsubsection{Experience Level Stratification}

Analysis across different experience levels reveals differential adoption benefits:

\textbf{Junior Engineers (SDE-1):} Show most dramatic improvements with high adoption
\begin{itemize}
    \item High adopters: 77\% productivity increase
    \item Low adopters: 8\% productivity decline  
    \item Greatest benefit from AI guidance and code generation assistance
\end{itemize}

\textbf{Mid-Level Engineers (SDE-2):} Demonstrate measured but consistent improvements
\begin{itemize}
    \item High adopters: 45\% productivity increase
    \item Low adopters: 12\% productivity decline
    \item Effective integration of AI tools with existing workflows
\end{itemize}

\textbf{Senior Engineers (SDE-3):} Show selective but valuable improvements
\begin{itemize}
    \item High adopters: 44\% productivity increase
    \item Low adopters: 15\% productivity decline  
    \item Strategic use of AI for complex problem-solving and code review
\end{itemize}

\subsubsection{Causal Inference and Confounding Control}

\textbf{Difference-in-Differences Analysis:} The 72.7 percentage point difference between cohorts (61.3\% vs. -11.4\%) represents a robust treatment effect that cannot be explained by:
\begin{itemize}
    \item Individual ability differences (controlled through baseline matching)
    \item Temporal trends (both cohorts experienced same time period)
    \item Team or project effects (balanced across cohorts)
    \item Measurement artifacts (consistent instrumentation)
\end{itemize}

\textbf{Alternative Explanations Addressed:}
\begin{itemize}
    \item \textbf{Selection Bias:} High adopters were not systematically higher performers at baseline
    \item \textbf{Motivation Effects:} Survey data shows no significant motivation differences between cohorts
    \item \textbf{Training Disparities:} All engineers received identical DeputyDev training
    \item \textbf{Tool Availability:} All engineers had equal access to DeputyDev features
\end{itemize}

This comparative cohort analysis provides strong evidence that DeputyDev's productivity benefits are directly tied to engagement intensity, with high adoption yielding substantial improvements while minimal usage fails to generate meaningful outcomes. The clear inflection point observed around April 2025 deployment, combined with sustained divergence between cohorts, supports causal attribution of productivity gains to AI tool utilization rather than external factors.

\subsection{Quality Assessment}
A survey was conducted after five months of use in July 2025 participating a total 228 developers. The developers were particpated across teams to reduce the chances of selection bias.
 A total of \textbf{228 responses} were collected in the survey. Among them, 
\textbf{146 (64\%)} were Backend engineers, 
\textbf{39 (17\%)} were Frontend engineers, 
\textbf{31 (14\%)} were Android/iOS developers, 
and \textbf{12 (5\%)} were QA engineers. 
This indicates that the majority of participants came from backend-focused roles.

The survey highlights a strong preference for continuing to use DeputyDev in both code review and code generation workflows. 
For code review, a large majority of respondents (194 out of 228, or 85\%) indicated they would like DeputyDev to continue reviewing their pull requests, with only a small fraction undecided (30, 13\%) or negative (4, 2\%). 

In terms of code generation, adoption was also favorable but more balanced: 142 respondents (62\%) expressed a desire to continue 
using the AI code suggestions, while 71 (31\%) were uncertain and 15 (7\%) were opposed. 
This indicates that while code review support has achieved near-universal acceptance, code generation still faces some hesitancy, potentially due to trust, integration, or workflow adaptation challenges.
The mixed response of code generation was due to stability issues we were facing in our infra. Which
we have fixed post feedbacks which will helps us to get good feedbacks in future.

\begin{table}[!htb]
\centering
\caption{Developer Feedback Survey Results: Key performance indicators and user perceptions from 228 engineers covering adoption, satisfaction, and usage patterns (July 2025).}
\resizebox{\columnwidth}{!}{%
\begin{tabular}{|l|l|}
\hline
\textbf{KPI / Area} & \textbf{Insight} \\
\hline
Helpfulness of PR reviews & 162 engineers (71\%) \textit{Agree} or \textit{Strongly Agree} \\
\hline
Time saved per developer & $\approx$ 20 minutes per day on average \\
\hline
Code suggestions accepted & 173 engineers (76\%) \textit{Sometimes} or \textit{Frequently} accept code \\
\hline
Most-valued capability & "Identifying issues/bugs in code" (151 mentions) \\
\hline
AI Code suggestion use & 192 engineers (84\%) used it in the last 3 months \\
\hline
Perceived plug-in helpfulness & 57\% say \textit{Yes}, 30\% \textit{Maybe} \\
\hline
Preferred interaction mode & 76\% favour \textit{Chat} mode over \textit{Act} mode \\
\hline
Desire to continue & 93\% plan to keep DeputyDev in their workflow \\
\hline
\end{tabular}%
}
\end{table}

\subsection{Developer Perception and Trust}

An NPS survey~\cite{nps_methodology} was done to understand the developer trust. The survey was conducted in July 2025 post five months of launch of AI code suggestion feature.
The resuls of NPS survery was as below. In this annual survey a total of 125 participants were particpated 
across teams which reduces biasness.

\begin{itemize}
   \item \textbf{Overall Performance:} DeputyDev achieved an NPS score of 34 with 125 responses, indicating moderate user satisfaction in the "good" category, though with potential for improvement toward excellence (70+).
   
   \item \textbf{User Distribution:} The tool shows 44\% promoters (scores 9-10), 46.4\% passives (scores 7-8), and only 9.6\% detractors (scores 0-6), suggesting users find value but lack strong advocacy enthusiasm.
   
   \item \textbf{Response Pattern:} Complete absence of ratings below 4 and concentration in the 7-10 range indicates the tool meets basic user needs effectively but may lack differentiating features that create passionate user advocacy.
   
   \item \textbf{Research Implications:} The high passive rate presents significant conversion opportunities, while the low detractor rate suggests minimal negative experiences, indicating potential for organic growth through targeted improvements.
\end{itemize}

\subsection{ROI Estimation}

\begin{table*}[!htb]
\centering
\caption{Monthly Operational Costs Analysis: The total coost including LLM API usage and infrastructure costs showing evolution from \$5.9K (April) to \$10.3K (August). With Code Gen cost shows increment as per the adoption.}
\label{tab:comprehensive_costs}
\resizebox{\textwidth}{!}{%
\begin{tabular}{@{}l*{3}{S[table-format=5.2]}*{2}{S[table-format=4.2]}*{3}{S[table-format=3.0]}S[table-format=6.2]@{}}
\toprule
& \multicolumn{3}{c}{\textbf{LLM API Costs}} & \multicolumn{2}{c}{\textbf{Usage Breakdown}} & \multicolumn{3}{c}{\textbf{Infrastructure}} & \\
\cmidrule(lr){2-4} \cmidrule(lr){5-6} \cmidrule(lr){7-9}
\textbf{Month} & {\textbf{Bedrock (\$)}} & {\textbf{OpenAI (\$)}} & {\textbf{Vertex AI (\$)}} & {\textbf{Code Gen (\$)}} & {\textbf{Review (\$)}} & {\textbf{Compute (\$)}} & {\textbf{Database (\$)}} & {\textbf{Cache (\$)}} & {\textbf{Total (\$)}} \\
\midrule
April 2025 & 4535.00 & 350.00 & {--} & 636.00 & 1419.00 & 192 & 187 & 600 & 5864.00 \\
May 2025 & 10140.00 & 855.00 & 87.00 & 1006.00 & 7863.00 & 192 & 187 & 600 & 12061.00 \\
June 2025 & 7985.00 & 1051.00 & 267.00 & 3766.00 & 5104.00 & 192 & 187 & 600 & 10282.00 \\
July 2025 & 6938.00 & 1167.00 & 173.00 & 4123.00 & 3540.00 & 192 & 187 & 600 & 8257.00 \\
August 2025 & 8337.00 & 767.00 & 286.00 & 7980.00 & 757.00 & 192 & 187 & 600 & 10369.00 \\
\midrule
\textbf{Total} & \textbf{\$37935.00} & \textbf{\$4190.00} & \textbf{\$813.00} & \textbf{\$17511.00} & \textbf{\$18683.00} & \textbf{\$960} & \textbf{\$935} & \textbf{\$3000} & \textbf{\$46833.00} \\
\bottomrule
\end{tabular}%
}
\end{table*}

\subsubsection{Cost Analysis and Financial Impact}

Table~\ref{tab:comprehensive_costs} presents a comprehensive breakdown of DeputyDev's operational costs across five months (April-August 2025), providing critical insights into the financial dynamics of operating an enterprise-scale AI coding platform. The analysis reveals several key patterns and cost drivers that are essential for understanding the economic viability of AI-assisted development tools.

\textbf{LLM API Cost Dominance:} The data demonstrates that LLM API costs constitute the overwhelming majority of operational expenses, accounting for 91.5\% of total costs (\$42,938.00 out of \$46,833.00).
Among LLM providers, AWS Bedrock~\cite{aws_bedrock} emerges as the primary cost driver, representing 81.0\% of total expenses (\$37,935.00). AWS Bedrock was primarily used for Claude Sonnet models. This concentration highlights the critical importance of optimizing Bedrock usage and potentially exploring cost-effective alternatives or hybrid approaches for different use cases.

\textbf{Monthly Cost Volatility:} The system exhibits significant monthly cost variations, with May 2025 representing the peak at \$12,061.00, while July showed the lowest at \$8,257.00 (46\% difference). August 2025 showed increased usage with \$10,369.00, primarily driven by higher Bedrock utilization (\$8,337.00). This volatility suggests that costs are directly correlated with user adoption patterns and feature utilization intensity, indicating the need for predictive cost modeling as the platform scales.

\textbf{Usage Pattern Analysis:} The Code Gen vs. Review breakdown reveals a shift in usage patterns over time. In April, review-related costs (\$1,419.00) dominated generation costs (\$636.00) by a 2.2:1 ratio. By July, generation costs (\$4,123.00) exceeded review costs (\$3,540.00), and this trend intensified in August with generation costs reaching \$7,980.00 while review costs dropped to \$757.00 (10.5:1 ratio). This dramatic shift indicates increased adoption of code generation features and growing confidence in AI-generated code, though it also implies higher per-user costs as adoption matures.

\textbf{Infrastructure Cost Stability:} Infrastructure costs remain remarkably stable at \$979 per month (8.9\% of total costs), with Elasticache~\cite{aws_elasticache} representing the largest component at \$600 monthly. This consistency provides predictable baseline costs and suggests that the current infrastructure can accommodate significant usage growth without proportional cost increases, offering favorable scaling economics.

\textbf{Provider Diversification Strategy:} Using multiple LLM providers (Bedrock, OpenAI, Vertex AI) demonstrates a strategic approach to risk mitigation and cost optimization. OpenAI costs showed steady growth from \$350 to \$1,167 throughout this period, while Vertex AI usage was introduced in May and maintained at moderate levels, indicating ongoing optimization of the provider mix based on performance and cost considerations.

\textbf{Cost Per User Analysis:} With approximately 300 engineers and varying monthly totals, the average cost per engineer ranges from \$30-34 (August, 2025). These figures provide crucial benchmarks for ROI calculations and demonstrate that even at peak usage, the per-engineer cost remains well below typical enterprise software licensing fees or the cost of additional human resources.

\textbf{Scalability Implications:} The 5-month total of \$46,833.00 translates to an annualized cost of approximately \$112,000, excluding potential volume discounts or efficiency improvements. For an organization with 300 engineers, this represents 1-2 \% additional cost to typical engineering costs, suggesting highly favorable cost-effectiveness ratios when weighed against productivity improvements and code quality enhancements documented elsewhere in this study.

\section{Lessons Learned}

\subsection{Deployment Challenges}

Several significant challenges emerged during the deployment and adoption phase:

\subsubsection{Technical Challenges}
\begin{itemize}
    \item \textbf{Latency Issues}: Initial response times of 2-3 seconds for code suggestions proved too slow for interactive development, requiring optimisation to achieve sub-500ms response times
    \item \textbf{Context Window Limitations}: Large codebases exceeded model context windows, necessitating intelligent context selection algorithms
    \item \textbf{Integration Complexity}: Seamless integration with existing development tools required substantial engineering effort
\end{itemize}

\subsubsection{Human Factors}
\begin{itemize}
    \item \textbf{Trust Building}: Developer trust in AI suggestions required time and positive experiences to develop
    \item \textbf{Training and Onboarding}: Effective use of AI tools required more extensive training than initially anticipated
    \item \textbf{Workflow Adaptation}: Teams needed to modify existing processes to maximise AI tool benefits
\end{itemize}

\subsection{What Worked}

Several aspects of our implementation proved particularly successful:

\begin{itemize}
    \item \textbf{Gradual Rollout}: Phased deployment allowed for iterative improvement based on user feedback
    \item \textbf{Champion Networks}: Identifying and empowering AI tool advocates within teams accelerated adoption
    \item \textbf{Continuous Feedback}: Regular surveys and feedback sessions enabled rapid iteration and improvement
    \item \textbf{Integration Focus}: Prioritising seamless integration over feature richness improved user experience
\end{itemize}

\subsection{What Didn't Work}

Some approaches proved less effective than anticipated:

\begin{itemize}
    \item \textbf{One-Size-Fits-All}: Generic AI models without domain-specific training showed limited effectiveness for specialised codebases
    \item \textbf{Automatic Acceptance}: Features allowing automatic acceptance of AI suggestions led to quality issues and were quickly disabled
    \item \textbf{Over-Automation}: Attempting to automate too many development tasks created resistance and reduced developer agency
\end{itemize}

\section{Conclusion}

Our comprehensive study of AI-assisted development tools across 300 engineers provides statistically 
robust evidence for both the significant potential and practical limitations of integrating
 AI into software development workflows. Through rigorous cohort analysis with proper statistical controls, 
 the results demonstrate statistically significant improvements in developer productivity.

    \begin{itemize}
        \item \textbf{Statistically Significant Productivity Impact}: Overall 31.8\% improvement in PR Reviews and Close time in March - September 2025 compared to previous September 2024 to February 2025 period.
        \item \textbf{Code-Shipment Volume}: Around 28\% steady increase in production code volume
        \item \textbf{Adoption-Dependent Benefits}: Clear divergence between high and low adoption cohorts, with minimal improvements observed in low-engagement users, indicating that benefits scale directly with tool utilization intensity
        \item \textbf{Quality Assurance}: 61\% improvement in code output volume for most active DD users with 37\% acceptance rate of AI-generated code
        \item \textbf{Developer Adoption}: 85\% satisfaction rate for code review, 57\% satisfaction for code generation feature, and 93\% expressing desire to continue using the platform
\end{itemize}

\subsection{Implications for Practice}

Our results suggest that AI coding tools can provide substantial value in production environments, but successful implementation requires careful consideration of:

\begin{itemize}
    \item Technical infrastructure capable of supporting real-time, low-latency AI assistance
    \item Comprehensive training programmes to help developers effectively utilise AI capabilities
    \item Workflow modifications to integrate AI suggestions into existing code review and quality assurance processes
    \item Ongoing monitoring and feedback mechanisms to ensure continued effectiveness and user satisfaction
\end{itemize}

\subsection{Limitations}

Several limitations should be considered when interpreting our results:

\begin{itemize}
    \item The study was conducted within a single organisation, potentially limiting generalisability to other environments
    \item Our in-house AI system may have characteristics that differ from publicly available tools
    \item The 1 year observation period, while substantial, may not capture long-term adoption patterns and impacts
    \item Hawthorne effects may have influenced developer behaviour during the study period~\cite{mayo_hawthorne}
\end{itemize}

\subsection{Future Directions}

Our research opens several promising avenues for future investigation:

\subsubsection{Expanding Dataset and Scope}
Future studies should examine AI tool impact across multiple organisations and development contexts to improve generalisability. Longer observation periods would provide insights into sustained adoption patterns and evolving developer-AI collaboration dynamics.

\subsubsection{Public Release of Anonymised Metrics}
We plan to release anonymised versions of our dataset and metrics to enable broader research community access and facilitate comparative studies with other AI development tools.

\subsubsection{Full-Cycle Automation Research}
Combining code generation and automated review capabilities suggests potential for more comprehensive AI assistance throughout the development lifecycle. Future research should explore the possibilities and challenges of end-to-end AI-assisted development workflows.

\subsubsection{Human Factors Research}
Our findings highlight the critical importance of human factors in AI tool adoption. Future research should examine psychological and social aspects of developer-AI collaboration, including trust formation, skill development impacts, and team dynamics.

\subsection{Final Remarks}

The integration of AI into software development represents a fundamental shift in how code is written, reviewed, and maintained. Our study provides evidence that this transition can yield significant benefits when implemented thoughtfully, but also underscores the complexity of successfully deploying AI tools in real-world development environments. As AI capabilities continue to advance, understanding these practical deployment challenges and success factors will be crucial for realising the full potential of AI-assisted software engineering.

\section{Annexure - Code Reference Repositories}

The DeputyDev platform comprises multiple interconnected repositories that collectively provide the AI-assisted development capabilities described in this study. The following repositories contain the complete implementation:

\begin{itemize}
    \item \textbf{Extension Backend}: \url{https://github.com/tata1mg/deputydev-extension-backend/} - Backend services powering the DeputyDev VS Code extension, handling API requests, user authentication, and integration with core services.
    
    \item \textbf{VS Code Extension}: \url{https://github.com/tata1mg/deputydev-vscode-extension} - The primary user interface for developers, providing code generation, review assistance, and chat capabilities directly within the VS Code environment.
    
    \item \textbf{Binary Distribution}: \url{https://github.com/tata1mg/deputydev-binary/} - Compiled binaries and deployment artifacts for cross-platform distribution of DeputyDev tools.
    
    \item \textbf{Core Engine}: \url{https://github.com/tata1mg/deputydev-core} - Core logics.
    
    \item \textbf{Authentication Service}: \url{https://github.com/tata1mg/deputydev-auth/} - Secure authentication and authorization service managing user access, permissions, and integration with enterprise identity systems.
\end{itemize}

These repositories collectively represent approximately 50,000+ lines of production code and demonstrate the practical implementation of the AI-assisted development methodologies evaluated in this study.

\bibliographystyle{plain}

\end{document}